# Symmetrical Dispersion Compensation For High Speed Optical Links

Ojuswini Arora[1], Dr.Amit kumar Garg[2], Savita Punia[3]

[1]Electronics & Communication Department

M.M.E.C, M.M University, Mullana, Ambala, 133203 Haryana, India

[2]Electronics & Communication Department

M.M.E.C, M.M.University, Mullana, Ambala, 133203 Haryana, India

[3]Electrical Department

M.M.E.C, M.M University, Mullana, Ambala, 133203 Haryana, India

**Abstract**

In this paper, the performance of high speed optical fiber based network is analysed by using dispersion compensating module (DCM). The optimal operating condition of the DCM is obtained by considering dispersion management configurations for the symmetrical system i.e Pre-compensation & Post-compensation. The dispersion compensating fiber (DCF) is tested for a single span, single channel system operating at a speed of 10 Gb/s with a transmitting wavelength of 1550 nm, over 120 km single mode fibre by using the compensating fiber for 24 km,30km and 35Km. So far, most of the investigations for single mode fiber (SMF) transmission at high amplifier spacings in the order of 90 km to 120 km is focused on conventional Non Return to Zero(NRZ) format. The simulation results are validated by analysing the Q-factor and Bit error rate (BER) in the numerical simulator OptSim.

**Keywords**: *Dispersion, Dispersion Compensation, Symmetrical compensation, Bit error rate (BER), Dispersion compensating fiber (DCF) , single mode fiber(SMF)*.

## 1. Introduction & Previous Work

It is seen that attenuation in the single-mode fiber is lowest in the 1.55- mm wavelength region (0.2 dB/km). However, the intramodal dispersion in this wavelength region is so severe that its effect is intolerable for very high *BL* systems. Because of the high dispersion the operating speed of the network reduces to 2.5 Gb/s. In order to take the advantage of lowest attenuation at this wavelength window, it was realized that the optical fibers need to be modified to have low dispersion in this window. In this technique, fibres of negative dispersion coefficient are made to alternate along the length of the optical link. Negative dispersion fibers (NDF) have a large dispersion in comparison to standard SMF, thus a relatively short NDF can compensate for dispersion accumulated over long links of SMFs. NDFs are easy to install and require little modification to an already existing system. The major disadvantage of NDF is that it exhibits a large attenuation in signal power, which results into using more optical amplifiers to be employed in the system. This in turn will overcome the other limitations in the system because the non-linear attributes of this fibre are considerably higher. So, which has been validated by using Symmetric Compensation (Pre, Post compensation).Also the results have been validated by numerical simulations with the optical simulator OptSim. Nutys et al.[1] investigated theoretically and experimentally the transmission performance of a 10 Gb/s repeater transmission system using DCF. The system configuration that was considered is a 360 km standard (1300 nm zero-dispersion) fiber transmission system with an optical repeater including DCFs located every 120 km (or every 2100 ps/nm dispersion). The transmitter was a DFB laser externally modulated by a zero-chirp LiNbO3 modulator with NRZ (non-return to zero), 440 PRBS data. The results of this investigation clearly demonstrate that the use of DCFs is an extremely effective method to overcome the chromatic dispersion in high-speed transmission systems. Weinert et al. [2] investigated the possibilities of 40 and 440 Gb/s time division multiplexing/wavelength division multiplexing (TDM/WDM) return-to-zero (RZ) transmission over embedded standard single-mode fibers (SMF) at a transmission wavelength of 1.55μ m both experimentally and theoretically. Dispersion of the SMF was compensated by a dispersion compensating fiber (DCF). Transmission over a span of 150 km of SMF in the single channel case and of 100 km SMF in the multichannel case is reported. It was shown numerically that improvement was achieved by employing the newest type DCF which also compensates the dispersion slope of the SMF. Mob et al.[3] theoretically and experimentally analyses advantages of nonlinear RZ over NRZ on 10 Gb/s single-span links. In India, during the past few years' fiber dispersion and nonlinearities have been studied and their impact on the system performance has been investigated. Sharma et al. [4] reviewed the various fiber dispersion compensation methods and investigated techniques for compensation of dispersion by differential delay method including the impact of higher order dispersion terms Kaler et al. [5] discussed the limitations due to Group velocity dispersion (GVD) on transmission distance, bit rate and laser line width including the higher order dispersion effects. The power penalty analysis for different realistic weight functions for combating the pulse broadening effects of group-velocity dispersion in a fiber-optic communication link using differential time delay method with higher-order dispersion terms [6] was discussed. Further, the propagation of signal and noise in the transmission medium to observe the validity of higher order dispersion terms [7] was described. The comparison of pre-compensation, post-compensation and symmetrical-dispersion compensation schemes for 10 Gb/s NRZ links using standard and dispersion compensated fibers was also investigated[8].

Section 2 discusses the proposed work regarding the DCF system configuration and its Symmetric Compensation taking into account the pre-compensation & post-compensation. Results for the simulation are validated and discussed in section 3, followed by the concluding remarks related to the dispersion compensation of various configurations in the Section 4.

## 2. Proposed Work

- The fiber based method employs the dispersion compensation through a small section of fiber length. There are various techniques such as dispersion compensation fiber (DCF), reverse dispersion fiber, negative dispersion fiber to compensate the dispersion of the system.

- Dispersion compensating fiber (DCF) is the predominant technology for dispersion compensation. It consists of an optical fiber that has a special design such as providing a large negative dispersion coefficient while the dispersion of the transport fiber is positive. A proper length of DCF allows the compensation of the chromatic dispersion accumulated over a given length of the transport fiber, although standard modules with predetermined dispersion values (with a typical granularity corresponding to the dispersion of 20 km of SSMF) are commercially available.

- The main advantage of this technology is the fact that it provides a broadband operation with a smooth dispersion property and good optical characteristics. In the first generation of DCF, only about 60% of the SSMF dispersion slope was allowed to be compensated. Now,100% slope-matching for both SSMF and E-LEAF is commercially available.

- However, dispersion compensation modules based on a first-generation DCF are largely deployed and their associated slope-mismatch is a problem we have to live with. DCF also presents a quite large insertion loss although improvements have been reported recently. Dispersion compensation modules based on DCF are also bulky and again, size reduction is expected in the future as bend loss reduction could allow a significant improvement in the compactness. The disadvantages of the fiber-based methods are the extra fiber loss, high non-linearities and an additional cost of the DCF is imposed. The maximum dispersion of such DCF is about -100 ps/nm/km, which is limited by the mismatching of the glass properties between the core and the cladding.

- Very long lengths of dispersion compensating fibers are required to compensate for the dispersion of even modest lengths of transmission. So, it is proposed by using the Symmetrical optimization for the optical fibers using Pre & Post compensation both. Dispersion management can be achieved with various combinations of fibre layout. The widely implemented configurations are Pre-compensation, Post-Compensation and Symmetrical compensation which uses both the techniques in one link as shown in Fig 1.

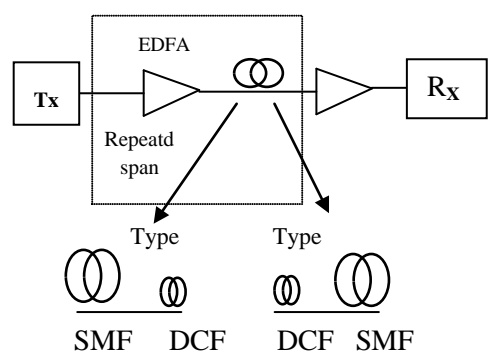

Fig.1 Dispersion management using Symmetric Compensation.

- To achieve optimum dispersion compensation, a dispersion management scheme is modelled using a transmission line consists of equal numbers of 120 km SMF and DCF sections of 24 km, 30 km and 35km for Pre compensation & Post compensation both.

- By varying the lengths of the Post compensating & the Pre compensating fibers, the fiber link is then tested out for an optimum compensation scheme, at which a minimum amount of dispersion, good eye opening, a.high quality factor ( Q- Factor) and a low BER is observed

Table1: Simulation Parameters used for the symmetrical compensation

| Simulation Parameters | SMF | DCF pre-compensation | DCF post-compensation |
|---|---|---|---|
| Length(km) | 120 | 24 | 24 |
| Dispersion [ps/km/nm] | 16 | -80 | -80 |
| Loss(dB/km) | 0.2 | 0.6 | 0.6 |
| Nonlinear Coefficient (Wkm)$^{-1}$ | 1.26677 | 1.8 | 1.8 |
| Transmitter & Receiver Parameters | NRZ | - | - |
| Bit rate [Gb/s] | 10 | - | - |

- The transmitter section consists of data source, electrical driver (NRZ), laser source (CW Lorentzian) and amplitude modulator (sin² MZ). The fiber parameters for SMF and DCF are listed in Table 1 and the simulated model is shown in fig.2. The figure comprises of equal sections of SMF & DCF ( Pre compensation & Post Compensation).

- In this technique, a partial compensation of second-order dispersion by employing DCF units is employed. In the zero path-average dispersion in all schemes or NRZ modulation formats, the transmitter emits chirp-free modulated pulses.

- Dispersion Compensation is employed to increase the transmission distance in systems operating at higher bit rates.Furthermore, Dispersion compensating devices are required to have a sufficiently large badwidth in order to achieve simultaneous across all the channels.

- Several dispersion and dispersion slope compensating devices are also available including single mode and higher order mode dispersion compensating devices. Although they have great potential for dispersion compensation but DCF are widely employed

- However fiber nonlinearity coefficient for DCF is 1.8 and for SMF is 1.266. At the receiver, the signal is electrically filtered by using Bessel filter(low pass), bandwidth,8dB. As a measure of system performance Q factor and BER are evaluated from the simulations in standard fiber transmissions operating at 10Gb/s at high amplifier spacings of I20km, the impact of fiber nonlinearity is diminished by symmetrical ordering of dispersion compensating fibres.

## 3. Results & Discussions

By taking into account the dispersion management scheme which is empoyled using a transmission line consisting of equal numbers of 120 km SMF and DCF sections of 24 km,30km and 35km for Pre compensation & Post compensation. The observations made are as shown in Table 2.

Table2: Simulation Results for the symmetrical compensation for various lengths of Pre & Post compensation

| Pre compensating Fiber length(Km) | Post compensating Fiber length(Km) | Q Value (dB) | BER |
|---|---|---|---|
| 24 | 24 | 30.76 | 1e-040 |
| 30 | 24 | 20.53 | 1.80072e-0.26 |
| 30 | 30 | 15.814 | 3.46214 e-010 |
| 35 | 35 | 7.1682 | .0111766 |

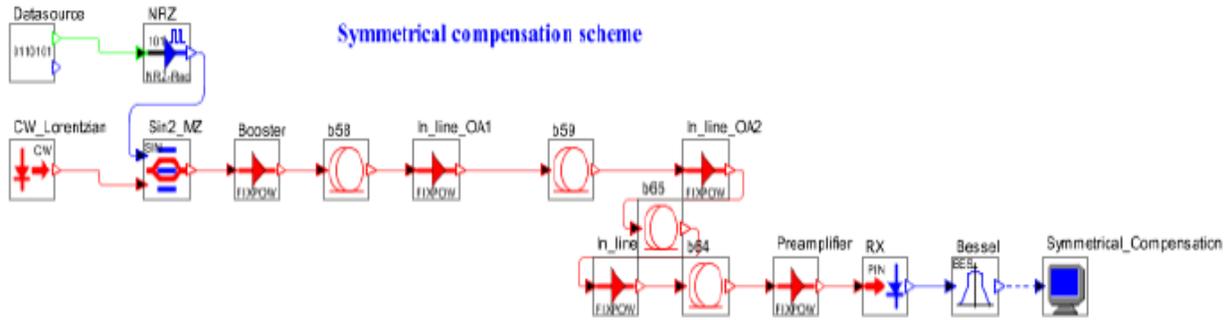

Fig.2 Simulated model for the Symmetrical Dispersion Compensation

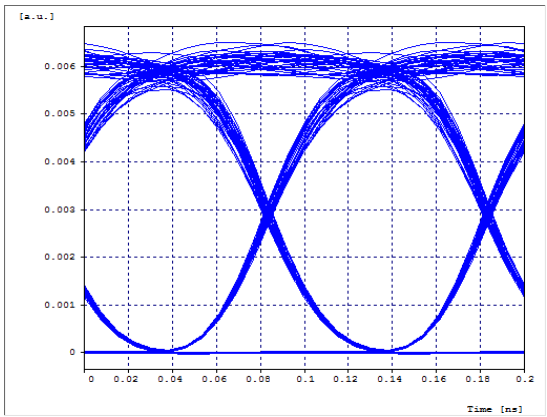

Fig.3 Pre-compensation fiber length=24km, SMF=120Km, Post-Compensation fiber length=24km,Q Value(dB)=30.76, BER=1e-0404

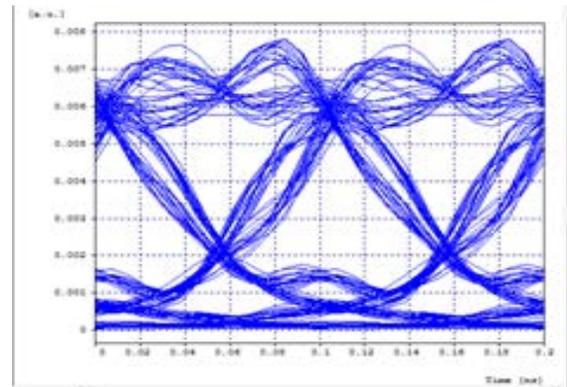

Fig.5 Pre-compensation fiber length=30km, SMF=120Km, Post-Compensation fiber length=30km.Q Value (dB)=15.814, BER=3.47e-01

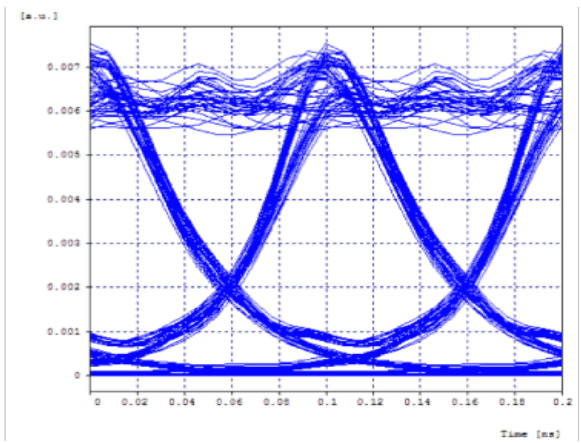

Fig.4 Pre-compensation fiber length=30km,SMF=120Km, Post-Compensation fiber length=24km, Q Value(dB)=20.53, BER=1.8e-0.26

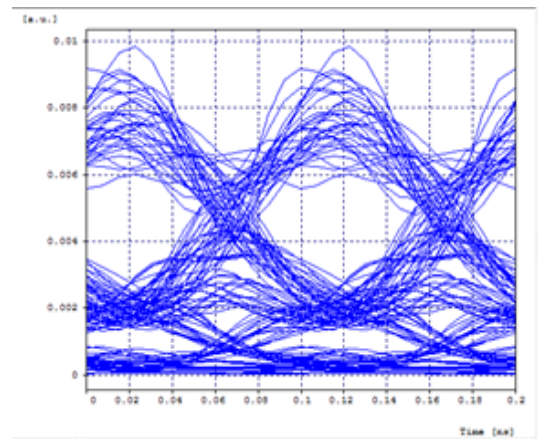

Fig.6 Pre- compensation fiber length=35km, SMF=120Km, Post- Compensation fiber length=35km,Q Value(dB) = 7.1682, BER=.01117

In the proposed methodology, a comparison is made between the various configurations taken, for pre-compensation & post-compensation. It is observed that the effect of dispersion compensating fibers using 30Km fiber for pre compensation, SMF of 120Km and 24Km fiber for the Post compensation, the estimated values of the Q Factor=20.5346dB, BER=1.80072 e 0.26 and Jitter=0.223014ns [fig.4]. Analysing other configurations of Pre-compensating fiber length=30km, SMF=120Km, Post-Compensation fiber length=30km [fig.5] and DCF Pre compensation fiber length=35km, SMF=120Km, Post-Compensation fiber length = 35km [fig.6]. However, if the DCF length for the Pre-compensation & Post compensation fiber = 24Km, SMF=120Km;the estimated values of the Q Factor=30.7602dB, BER=1e-040 and Jitter=0.024919ns [fig3], this is the optimum compensation scheme which has been obtained.

In this work, it is found that the system performance gradually improved as the total dispersion of the transmission fiber tend towards that of the DCF and in a similar fashion, the system performance decreases as the total dispersion of fiber exceeded that of the DCF. Results show that symmetrical compensation for 24Km fiber of DCF is much better than the results obtained for the compensation obtained for the 30Km and 35Km length of the fiber. Furthermore, analysis of the Q-factor revealed that system performance has exceeded by an amount of 10dB which shows a significant increase..

## 4. Conclusions

The work has emphasised on the dispersion compensation techniques at 10Gb/s. A simulation model is presented for the dispersion compensation in optical fibers for a single channel single span fiber. It is observed that the length of dispersion compensating fiber (DCF) effects the dispersion compensation for the schemes Pre-compensation & Post- compensation. The results show that the optimum compensation scheme is achieved when a DCF of 24km for pre & post compensation is used. It has been found that there is a considerable improvement in terms of Q factor by an amount of 10dB when the results are being compared with the other schemes used.

**Ojuswini Arora** B.Tech in Electronics & Instrumentation from Kurukshetra University, Pursuing M.TECH  in Electronics & Communication from M.M. university. She has published research articles in International Journals. Currently employed as a lecturer in Electrical Departmnet, M.M.E.C, M.M. University, Mullana, INDIA

**Dr. Amit Kumar Garg** Received his B.E. and M.E. degrees in Electronics & Communication Engineering with distinction from Gulbarga University and Panjab University (India) respectively., Ph.D from Thapar University, Patiala. he is  presently working as Professor  & Head in Electronics & Communication Engineering   Departmernt at M.M.E.C, M.M. University, He has published several scholarly articles in various international journals of high repute & is also the reviewer of various esteemed journals.

**Savita Punia** M.tech pursuing form Electrical Department, M.M..University, Mullana, .Currently employed in Electrical Department M.M.E.C, M.M.University, Mullana.